\documentclass[12pt]{iopart}
\usepackage{amssymb}
\usepackage{graphicx}
\usepackage{epsfig}

\begin{document}
\title[Doping dependence of $T_{\rm c}$ of superconducting boron-doped diamond]{Model for the boron-doping dependence of the critical temperature of superconducting boron-doped diamond }

\author{B\v{r}etislav \v{S}op\'{\i}k $^{1,\,2}$}
\address{$^1$ Faculty of Mathematics and Physics, Charles University, \\ Ke Karlovu 3, 12116 Prague 2, Czech Republic} \address{$^2$ Institute of Physics, Academy of Sciences, Cukrovarnick{\' a} 10, 16253 Prague 6, Czech Republic}
\ead{sopik@fzu.cz}

\begin{abstract}
We study the concentration dependence of the superconducting critical temperature $T_c$ in a boron-doped diamond. We evaluate the density of states at Fermi level $N_0$ within the dynamical cluster approximation obtaining higher values than from the coherent potential approximation. We discuss the $T_c$ as a function of $N_0$ within the BCS, the McMillan, and the Belitz theory. The simplified Belitz theory gives the best agreement with experimental data. Since the density of states follows a simple power-law for accessible doping concetrations $x$, the present theory offers an analytical formula for $T_{\rm c}(x)$.
\end{abstract}

\pacs{74.25.Dw, 71.23.An}
\submitto{\NJP}

\maketitle

\section{Introduction}

A diamod doped with boron can become superconducting at temperatures of few kelvins \cite{Ekimov}. Such critical temperatures are surprisingly high for impurity band conductivity. Moreover, one can expect that $T_{\rm c}$ will be soon further increased. It would be useful to predict an optimal sample composition from a microscopic theory. To this end we evaluate the density of states at the Fermi level within the dynamical cluster approximation and show that it can be approximated by a simple power-law. Using this density of states the $T_{\rm c}$ can be predicted from a McMillan type formula.

Diamond is an insulator with a band gap of $\sim 5.5 \,{\rm eV}$. Boron atoms create shallow acceptor levels close to the valence band with activation energy of holes $\sim 0.37\,{\rm eV}$. At a doping concentration above $\sim 4.5 \times 10^{21} \, { \rm cm}^{-3}$ the system presents a metal-insulator transition \cite{Klein} and might become superconducting at $0.5$~K. With increasing concentration $T_{\rm c}$ increases. A favoured theory of superconductivity driving mechanism is, that electrons mostly interact with localized vibrational modes on the boron atoms in an optic spectra. However other possible theories are discussed as well \cite{Mares}.

Experimental data on boron doped diamond are controversial. Majority of samples were prepared in thin layers and their properties strongly depend on crystallographic orientation of the surface. There are two widely studied types with $100$ and $111$ orientation, see references \cite{Klein} and \cite{Takano1}, respectively. Both groups have used the Microwave Plasma assisted Chemical Vapor Deposition method though under different growth conditions. Their samples differ in many characteristics including the $T_{\rm c}$. The $111$ samples have higher $T_{\rm c}$ in general. In this paper we focus on the $100$ samples which are also supported by bulk samples prepared with the High Temperature High Pressure method \cite{Ekimov} revealing a comparable $T_{\rm c}$.

The theory of boron-doped diamond can benefit from an extensive experience with superconductivity in disordered materials. \cite{Weinkauf, Wysokinski1, Wysokinski2, Belitz} Unlike in metallic alloys, where the disorder mainly modifies the coherence length and can be handled within simple approximations, in the case of doping induced superconductivity a sophisticated treatment of the disorder is necessary to describe a formation of the impurity band. Shirakawa \etal \cite{Shirakawa} have shown that a concentration dependence of $T_{\rm c}$ measured in boron doped diamond cannot be reproduced by the BCS theory in which the disorder is described on the level of the Coherent Potential Approximation (CPA). Since this theory yields too low values of $T_{\rm c}$, they proposed that disorder effects beyond the CPA are responsible for this disagreement. We will show that corrections beyond the CPA do increase $T_{\rm c}$, however in the extent which is not sufficent to cover discrepancies in question.

In the present paper we compute the density of states at the Fermi energy level for boron doped diamond within the Dynamical Cluster Approximation (DCA). While the CPA assumes a single atom embeded in the effective medium, the DCA generalises this idea by embedding a cluster. The cluster allows us to describe splitting of closely located impurity bound states which contributes to the shape of the impurity band. With respect to the superconductivity it is essential that the density of states at the Fermi level computed within the DCA is higher than the CPA value.

Apparently, the density of states at Fermi level is much lower in the impurity band than in ordinary metals. Since values of $T_{\rm c}$ are comparable, the boron doped diamond belongs to the family of the strong coupling materials. One can expect large dicrepancies from the BCS theory and we will show, that this is indeed the case. In this paper we use the Belitz theory \cite{Belitz} which generalises the McMillan formula to disordered superconductors.

The structure of the paper is as follows. In Sec. II we simplify the Belitz theory so that it depends on the doping exclusively via the disorder dependent density of states. In Sec. III we evaluate the density of states within the dynamical cluster approximation. The coherent potential approximation is evaluated for comparision. In Sec. IV we discuss the concentration dependence of the critical temperature comparing predictions of the Belitz theory, the McMillan formula and the BCS theory. To this end we first establish density independent material parameters with the help of which one can construct the density dependent coupling strength $\lambda$, the Coulomb pseudopotential $\mu$ and the Belitz correction $Y'$. Section V brings conclusions.

\section{Belitz Theory}

Let us outline the Belitz theory, first. In the absence of currents, the Eliashberg selfenergy has three terms
$
	\Sigma(\epsilon,\omega) = \big(1 - Z(\epsilon,\omega)\big)\omega \, \hat \tau_{0} + Y(\epsilon,\omega) \, \hat \tau_{3} + \phi (\epsilon,\omega) \, \hat \tau_{1} \,,
$
where $\epsilon$ is the kinetic energy and $\omega$ is the Matsubara frequency. For pure superconductors $\phi$ renormalised by $Z$ gives a gap function $\Delta$ while $Y$ is a scalar that only shifts the chemical potential having no effect on the critical temperature. In disordered superconductors, parameter $Y$ becomes energy and frequency dependent. Belitz has shown that its energy derivative at Fermi energy $
	Y' = \frac{\rmd \,}{\rmd \,\epsilon} Y(\epsilon,\omega) \big \vert_{\epsilon = 0}
$ modifies the McMillan formula as
\begin{equation}
	\label{Belitz}
	T_{\rm c} = \frac{\omega_{\rm D}}{1.45} \exp \Bigg \lbrack - \frac{1.04\,(1 + \lambda + Y')}{\lambda - \mu^{*}\big(1 + 0.62\,\lambda/(1 + Y')\big)} \Bigg \rbrack\,.
\end{equation}
Here $\lambda$ describes an electron-phonon coupling and 
\begin{equation}\label{mu_star}
	\mu^{*}  = \mu \left \lbrack 1 + \frac{\mu}{1 + Y'} \ln\frac{\omega_{\rm C}}{0.62 \,\omega_{\rm D}}\right \rbrack^{-1}
\end{equation}
is the screened pseudopotential, in which $\mu$ characterises a strength of the Coulomb interaction, $\omega_{\rm C}$ is its effective range and  $\omega_{\rm D}$ is the Debye frequency. Setting $Y' = 0$ one recovers the McMillan formula.

In general, Belitz formula \eref{Belitz} includes four disorder dependent material parameters $\omega_{\rm D}$, $\lambda$, $\mu$ and $Y'$. Based on ab-initio computations \cite{Boeri, Giustino} which show that $\omega_{\rm D}$ changes negligibly with doping we ignore the disorder dependence of $\omega_{\rm D}$ and use a pure diamond value $\hbar \omega_{\rm D}/k_{B} = 1860 {\rm K}$. The strong disorder dependence of remaining parameters stems from a rapid variation of the density of states at the Fermi energy level $N_{0}$ in vicinity of the metal-insulator transition. 

We will use a simplified version of the Belitz theory neglecting disorder corrections to interaction vertices in perturbation series. This brings a significant reduction of numerical demands, because $\lambda$, $\mu$ and $Y'$ reduce to functions of $N_{0}$ only. Moreover, with the vertex correction neglected,
\begin{equation}\label{N_Y}
	Y' = 2 \lambda - \mu \,.
\end{equation}
so that we are left with $\lambda$ and $\mu$ similarly as in the McMillan theory.

We take $\lambda$ from ab-initio computations. Since the published results cover only few concentrations, we evaluate $\lambda$ in the spirit of Morel and Anderson \cite{Morel} formula
\begin{equation} \label{N_lambda}
	\lambda = \frac{U N_{0}}{1 + Q N_{0}}\,.
\end{equation}
Here $U$ represents a phonon-electron coupling strength and $1 + Q N_{0}$ describes a screening. 

The pseudopotential $\mu$ reads \cite{Belitz,Vonsovsky}
\begin{equation} \label{preN_mu}
	\mu = V N_{0} \bigg \lbrack 1 + \frac{V N_{0}}{1 + Y'} \ln\frac{E_{\rm F}}{\omega_{\rm C}}\bigg \rbrack^{-1} \,,
\end{equation}
where $V$ is the strength of the Coulomb interaction. Since we have found no measurement of $V$ in the literature, we have to treat it as a parameter which is set from experimental data. Because in the boron-doped diamond the Fermi energy level is localised very close to the top of the valence band we associate $E_{\rm F} \approx \omega_{\rm C}$ with the effective range of the coulombic interaction and estimate both as the distance of the Fermi level from the top edge of the impurity band. The equation \eref{preN_mu} then has a simple form
\begin{equation}\label{N_mu}
 \mu = V N_{0}\,.
\end{equation}

Using formulas \eref{N_Y}, \eref{N_lambda} and \eref{N_mu} in \eref{Belitz} and \eref{mu_star}, we obtain $T_{\rm c}$ as a function of $x$.

\begin{figure}
	\centerline{\psfig{file=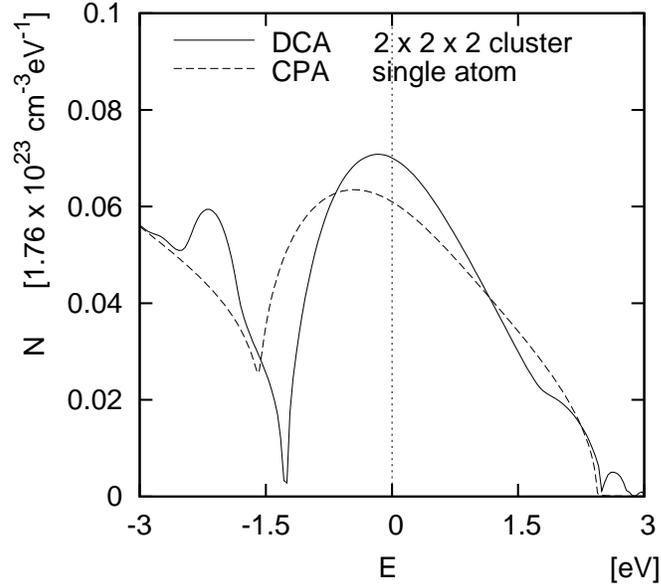,width=8cm,angle=-90}}
	\caption{\label{fig0_DOS} Impurity band of the density of states $N$ as a function of energy $E$ computed at $x = 0.05$ doping using the CPA and the DCA on the cluster of $2 \times 2 \times 2$ atoms. }
\end{figure}

\begin{figure}
	\centerline{\psfig{file=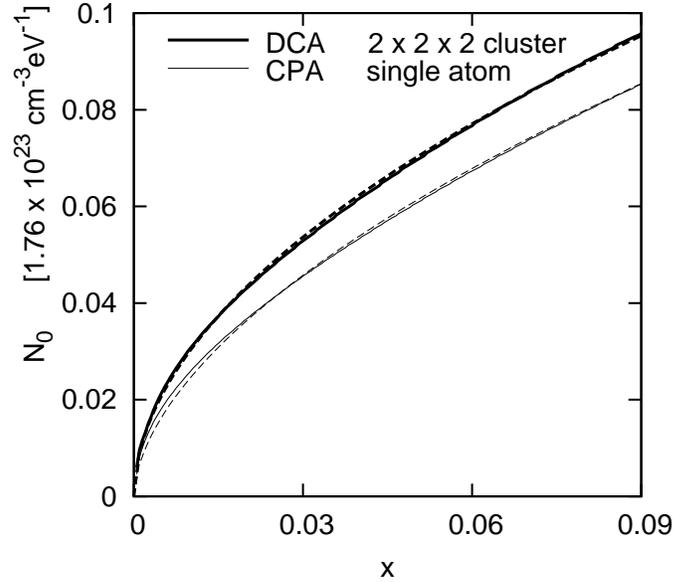,width=8cm,angle=-90}}
	\caption{\label{fig1_N} Density of states at the Fermi energy $N_{0}$ as a function of boron concentration $x$ computed using the CPA method and the DCA on the cluster of $2 \times 2 \times 2$ atoms. The figure includes dashed curves $N_{0}^{\rm CPA} \approx 0.59\,x^{0.568} \times 10^{23}\, {\rm cm}^{-3} {\rm eV}^{-1}$ approximating CPA and $N_{0}^{\rm DCA} \approx 0.59\,x^{0.523} \times 10^{23} \,{\rm cm}^{-3} {\rm eV}^{-1}$ for DCA. In most regions approximations are indistinguishable from computed results within the linewidths.}
\end{figure}

\section{Density of States}

To proceed we have to establish $N_{0}$ as a function of the boron concentration $x$. This single particle property is  independent of the electron-electron interaction, would it be the Coulomb repulsion or the phonon-mediated pairing potential. The density of states is thus given by a Hamiltonian of the valence band in diamond $H_0$ and a random potential of boron impurities 
\begin{equation}\label{hamilt}
\hat H = \hat H_0 + \sum_{i} \eta_{i}\delta \, \hat a^{\dagger}_{i} \hat a_{i}\,.
\end{equation}
In the random potential $\eta_i=1$ at impurity sites and zero elsewhere, and $\delta$ is the potential amplitude. We note that this single-site s-type potential does not cover the triple degeneracy of impurity state of boron. We adopt this Hamiltonian already studied by Shirakawa \etal \cite{Shirakawa} to make a link between their and present results.

We have evaluated $N_0$ within the DCA for clusters of $2 \times 2 \times 2$ atoms on a cubic lattice. Compared to the CPA density of states employed by Shirakawa \etal, our cluster density of states includes nontrivial corrections given by bonding and anti-bonding states at neighbour impurity sites. Difference between the CPA and DCA density of states in the impurity band is ilustrated in figure~\ref{fig0_DOS} for 5\%\ of boron doping. 

Within the CPA (dashed line) the impurity band yields rather featureless density of states having a slightly skewed semieliptic shape. The valence band starts approximately at an energy of $-1.5$~eV and ends at $-23$~eV. We focus on the impurity band because the Fermi energy, $E = 0$, lies there. Since boron is an acceptor, its bound states have higher energy than extended states in a valence band.

In the DCA (solid line) one can distinguish additional contributions of two and three boron clusters. Clusters of four to eight borons are also included, but their contributions are invisible for the given concentration. The main part of the impurity band density of states is formed by a bound state on a single boron. A shoulder on the right side of the impurity band results from a symmetric bound state of two neighbouring borons. The nonsymmetric state is not bounded but forms a resonant state visible in the valence band. The tree boron bound state is splited off at an energy $2.8$~eV. Higher states of the three boron clusters make a negligible contribution.

With respect to superconductivity the Fermi energy region is the most important. As one can see the density of states at $E = 0$ is dominated by the single boron states. This is a reason why the CPA and DCA give comparable $N_0$.

The resulting density of states at the Fermi level $N_0$ is presented in figure~\ref{fig1_N}. For comparison we also show the CPA result. One can see that the DCA density of states is higher than the CPA value at all impurity concentrations. Both resulting densities with a good accuracy obey power-law, $N_0^{{ \rm DCA}}\approx 0.59\,x^{0.568} \times 10^{23}\, {\rm cm}^{-3} {\rm eV}^{-1}$ and  $N_0^{{\rm CPA}}\approx 0.59\,x^{0.523} \times 10^{23}\, {\rm cm}^{-3} {\rm eV}^{-1}$ 
. A very good approximation by power-law also holds for the distance of the Fermi level from the top of the impurity band, $\omega_{\rm C} \approx 45.5\,{N_{0}}^{1.18}\, {\rm eV}$ for the DCA. In the rest of this section we provide more details of our computation. The reader not interested in these technical details can skip to the next section.

We only briefly introduce the DCA here, the reader can find all details in reference~\cite{Jarrell}. 
A basic idea of the dynamical cluster approximation is to divide the Brillouin zone into subzones -- in our case $2 \times 2 \times 2$ subzones. Within the subzone the selfenergy is momentum-independent, i.e., the selfenergy is represented by eight complex functions of frequency. By symmetry arguments this number can be reduced to four functions.

All subzones contribute to the density of states
\begin{equation}
	\rho^{0}(E) = \sum_{\textbf{K}} \rho^{0}_{\textbf{K}}(E) \,,
\label{denstat}
\end{equation}
where $\textbf{K}$ is a subzone index. In general, the subzone contribution $\rho^{0}_{\textbf{K}}$ is obtained integrating over the subzone with the electron dispersion of the valence band. For simplicity we approximate these functions 
by a semielliptical distribution with a proper width, shift and normalisation
\begin{equation}\label{semielipt}
	\rho^{(0)}_{\textbf{K}}(E) = \frac{1}{N_{\rm c}}\frac{2}{\pi \, u_{\textbf{K}}} \sqrt{1 - \bigg ( \frac{E - v_{\textbf{K}}}{u_{\textbf{K}}} \bigg )^{2} } \,,
\end{equation}
where $N_{\rm c}$ is number of subzones (8 in our case), and
\begin{eqnarray}
	u_{\textbf{K}} = 1/2 \big(E^{{\rm max}}_{\textbf{K}} - E^{{\rm min}}_{\textbf{K}}\big) \,, \\
	v_{\textbf{K}} = 1/2 \big(E^{{\rm max}}_{\textbf{K}} + E^{{\rm min}}_{\textbf{K}}\big) \,
\end{eqnarray}
are width and center of subzone energy bands with $E^{{\rm max}}_{\textbf{K}}$ and $E^{{\rm min}}_{\textbf{K}}$ being the maximum and minimum energy in the subzone $\textbf{K}$, respectively.

The density of states \eref{denstat} by definition maintains the width of the valence band. Since one of the subzones is centered around the $\Gamma$ point of the Brillouine zone, this approximation also yields a correct curvature at the edge, i.e., it correctly reproduces an effective mass of holes near the band edge. This feature is vital for a realistic description of the relatively shallow impurity state. We employ the model valence band \eref{denstat} in the CPA as well.

\begin{figure}
	\centerline{\psfig{file=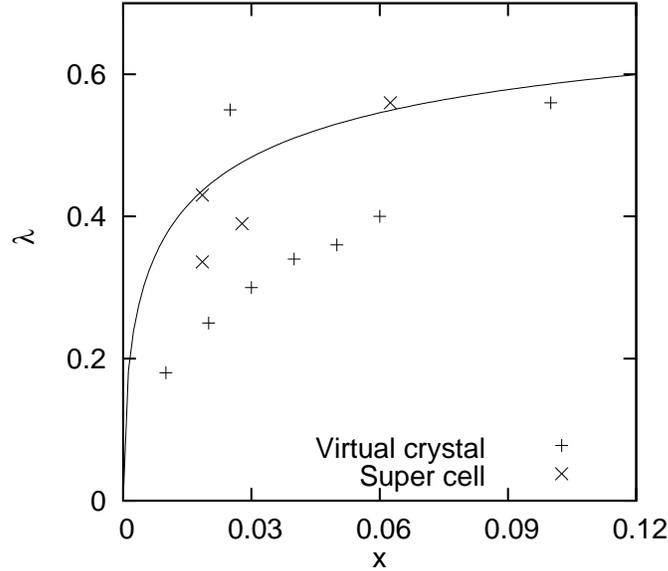,width=8cm,angle=-90}}
	\caption{\label{fig3_l} Electron-phonon coupling $\lambda$ as a function of boron concentration $x$. Crosses are results of ab-initio computations using super cell method \cite{Blase, Xiang, Giustino}. Plus-signs represent virtual crystal computations \cite{Lee,Boeri,Ma}. Solid line is Morel-Anderson formula \eref{N_lambda} with $U = 42.31  \times 10^{-23} \,{\rm cm}^{3} {\rm eV}$ and $Q = 54.65  \times 10^{-23} \,{\rm cm}^{3} {\rm eV}$. Densities of states have been associated to individual computations via DCA results shown in figure~\ref{fig1_N}.}
\end{figure}

\begin{figure}
	\centerline{\psfig{file=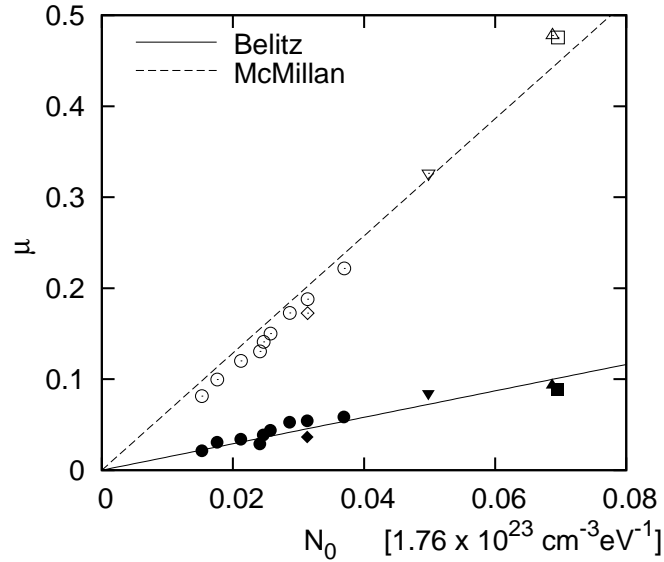,width=8cm,angle=-90}}
	\caption{\label{fig4_mu} Coulomb pseudopotential $\mu$ as a function of density of states $N_{0}$ of boron doped diamond. Full (empty) symbols are values of $\mu$ deduced by Belitz (McMillan) formula from experimental data of Ref.~ \cite{Klein}~$\opencircle$,  Ref.~\cite{Bustarret}~$\opendiamond$, Ref.~\cite{Ekimov}~$\opentriangledown$, Ref.~\cite{Takano3}~$\opentriangle$ and Ref.~\cite{Mukuda}~$\opensquare$. We employ $\lambda$ given by solid line in figure~\ref{fig3_l}. The full symbols fall close to the solid line of the theoretical pseudopotential \eref{N_mu} with the Coulomb strength $V = 2.54  \times 10^{-23} \,{\rm cm}^{3} {\rm eV}$ and the empty symbols are well reproduced with $V = 8.18 \times 10^{-23}\, {\rm cm}^{3} {\rm eV}$.}
\end{figure}

The parameters of the hamiltonian \eref{hamilt} are fitted according to the facts that the valence band is $22 \, {\rm eV}$ wide and the single impurity bound state appears at energy $0.37 \,{\rm eV}$ above the valence one. Using the local Green function corresponding the density of states \eref{denstat}, we come to $\delta = 8.91$~eV. 

We determine the Fermi energy level from a condition for a local averaged density of electrons at zero temperature $n_{\sigma} = \int_{\mu}^{\infty} \rho(E) \,dE  = x/2$. We are aware that a measured number of charge carriers can differ significantly from a number deduced from the concentration of boron atoms in the sample. As has recently been pointed out by Mukuda \etal \cite{Mukuda} quite a large fraction of boron atoms can form the neutral B-H complexes, reducing the concentration of charge carriers. On the other hand Klein \etal \cite{Klein} found that the effective number of carriers deduced from Hall-effect measurements was much larger than the number of boron atoms in samples. Since the charge carrier concentrations for all samples are not accessible we assume for simplicity the films to be doped ideally.

\section{Critical Temperature}

In this section we discuss the concentration dependence of the critical temperature $T_c$. To be able to implement the Belitz theory we need the coupling strength $\lambda$ and the pseudopotential $\mu$ as functions of $N_0$. 

The coupling strength $\lambda$ we deduce from ab-initio computations. In literature one finds the studies within the virtual crystal approximation \cite{Lee, Boeri, Ma} and the supercell method \cite{Blase, Xiang, Giustino}. 
In figure~\ref{fig3_l} we show fit of ab-initio results by formula \eref{N_lambda}. We found that parameters $U = 42.31 \times 10^{-23} \, {\rm cm}^{3} {\rm eV}$ and $Q = 54.65 \times 10^{-23} \, {\rm cm}^{3} {\rm eV}$ yield a reasonable fit of rather scattered computed values.

It remains to establish the pseudopotential. According to formula \eref{N_mu} we have to find a single parameter $V$, which determines $\mu$ for all concentrations. Dots in figure~\ref{fig4_mu} show values of $\mu$ deduced from the Belitz theory and experimental values of $T_c$. One can see that all dots stay close to a line given by formula \eref{N_mu} with $V=2.54 \times 10^{-23} {\rm cm}^{3} {\rm eV}$. The parameter $V$ holds for all concentrations and it is the only material parameter fitted to experimental values of $T_c$ in the present theory. 

\begin{figure}
	\centerline{\psfig{file=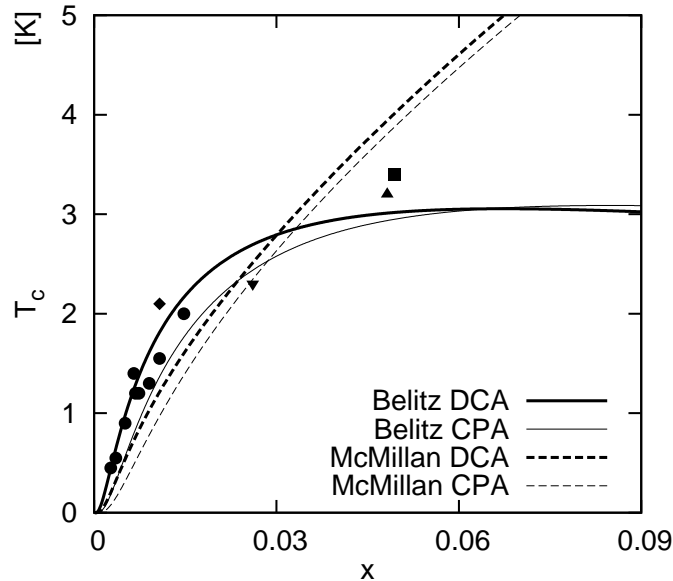,width=8cm,angle=-90}}
	\caption{\label{fig5_T} Critical temperature $T_{\rm c}$ as a function of boron concentration $x$. Symbols are data as in figure~\ref{fig4_mu}. Thick (thin) solid line is a result given by the Belitz theory using the DCA (CPA) density of states. Results of the McMillan theory are in dashed lines.}
\end{figure}

Now all relations and parameters are ready for predictions of the critical temperature. The concentration dependence of $T_c$ given by the Belitz formula \eref{Belitz} and power-law approximation of the DCA value of $N_0$ is shown in figure~\ref{fig5_T}. As one can see, the theory describes a steep increase of $T_c$ with doping at the region of small concentrations. At higher concentrations the critical tempeture satures at about 3 K. This saturation reminds trends described by the theory of Osofsky \etal \cite{Osofsky} based on a heuristic rescaling of the BCS parameters.

For comparison we also show the value predicted by the Belitz theory for the CPA density of states. As one expects from the lower density of states, the CPA leads to lower values of $T_{\rm c}$, namely in the region of small concentrations. Our results thus confirm a trend predicted by Shirakawa \etal that corrections beyond the CPA will lead to higher $T_{\rm c}$.

We would like to emphasis that the disorder corrections by Belitz are necessary for a good agreement between experimental data and theory. It can be seen from attempt to the fit $T_{\rm c}$ with the McMillan formula as shown in figure~\ref{fig5_T}. In the comparison we have not used the pseudopotential $\mu$ fitted within the Belitz theory, but made a seprate fit directly from the McMillan formula, see figure~\ref{fig4_mu}.

\begin{figure}
	\centerline{\psfig{file=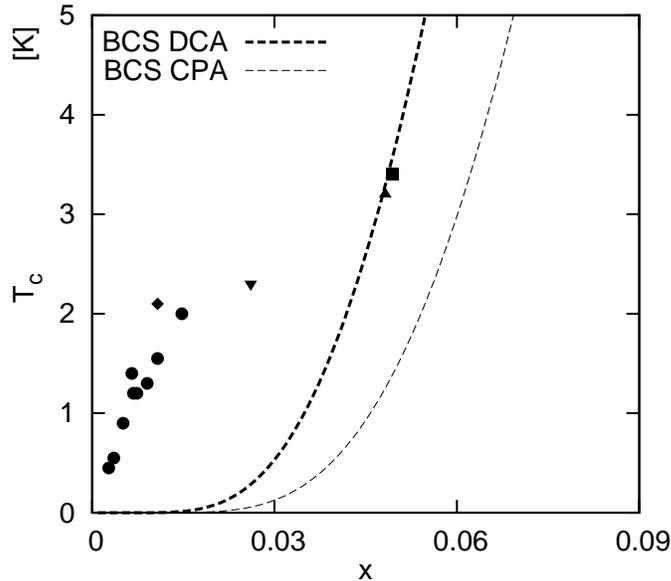,width=8cm,angle=-90}}
	\caption{\label{fig2_BCS} Failure of the BCS approach to the boron doped diamond. While experimental data (symbols) show decreasing slope of $T_{\rm c}$ as a function of boron concentration $x$, the BCS theory (dashed lines) predicts steeply increasing slope. The parameter $V_{\rm BCS} = 3.97  \times 10^{-23} \,{\rm cm}^{3} {\rm eV}$ was fitted within the DCA $N_{0}$ computations. Thick (thin) line is given by the DCA (CPA) $N_{0}$ results. Comparing the DCA (thick) and CPA (thin) one can see that the BCS theory overestimates effects of corrections beyond the CPA on the critical temperature.}
\end{figure}

Finally we show that the boron-doped diamond has to be treated as a strong coupling superconductor. In figure~\ref{fig2_BCS} we show $100$ experimental $T_{\rm c}$ data and its description with the BCS formula $T_{\rm c} = 1.14 \exp \left(-1/N_{0}V_{{\rm BCS}}  \right)\,$, where $V_{{\rm BCS}} = 3.97  \times 10^{-23} \,{\rm cm}^{3} {\rm eV}$ is the BCS interaction. Apparently, the BCS theory yields an incorrect concentration dependence. It is noteworthy how much the corrections beyond the CPA increase $T_c$ within the BCS theory. In the strong coupling theory these corrections are smaller by an order of magnitude.

\section{Conclusion}

To conclude we have discussed corrections beyond the CPA to the critical temperature in diamond doped with boron. From numerical studies of the dynamical cluster approximation with cluster of $2 \times 2 \times 2$ atoms it follows that these corrections increase the density of states at Fermi level. As expected this causes an increase of the critical temperature. Comparing different approximations we have shown that the strong coupling theory is necessary to predict realistic critical temperature. It also turns out that disorder corrections of Belitz improve agreement with experimental data. 

The present theory offers a simple analytic formula for the critical temperature as a function of boron density. Indeed, in a simplified Belitz theory all material parameters depend exclusively on the density of states and can be expressed as its linear or rational functions. Numerical studies of $C_{1-x}B_{x}$ show that the density of states at the Fermi level follows a power-law $N_{0} \approx 0.591\,x^{0.523} \times 10^{23}\, {\rm cm}^{-3} {\rm eV}^{-1}$, what allows one to predict the critical temperature without any numerical effort.

\begin{ack}
Author would like to thank Pavel Lipavsk{\' y} for many useful discussions and also to J. H. Samson for a critical reading of the manuscript. The access to the METACentrum computing facilities provided under the research intent MSM6383917201 is also highly appreciated. This work was supported by Grants No. GA{\v C}R 202/07/0597, No. AV0Z10100521, No. GAUK 135909, and No. GAAV 100100712, the DAAD and European ESF program AQDJJ.
\end{ack}

\end{document}